\documentclass[11pt,onecolumn,draftcls]{IEEEtran}
\def\comment#1{}
\usepackage{graphicx,subfigure,subfig,epsfig,amsfonts,amsmath,amssymb,lscape,color}
\usepackage[noadjust]{cite}
\DeclareSymbolFont{grb}{OML}{cmm}{b}{it}
\DeclareMathSymbol{\alphab}{\mathord}{grb}{"0B}
\DeclareMathSymbol{\betab}{\mathord}{grb}{"0C}
\DeclareMathSymbol{\gammab}{\mathord}{grb}{"0D}
\DeclareMathSymbol{\deltab}{\mathord}{grb}{"0E}
\DeclareMathSymbol{\epsilonb}{\mathord}{grb}{"0F}
\DeclareMathSymbol{\zetab}{\mathord}{grb}{"10}
\DeclareMathSymbol{\etab}{\mathord}{grb}{"11}
\DeclareMathSymbol{\thetab}{\mathord}{grb}{"12}
\DeclareMathSymbol{\kappab}{\mathord}{grb}{"14}
\DeclareMathSymbol{\lambdab}{\mathord}{grb}{"15}
\DeclareMathSymbol{\mub}{\mathord}{grb}{"16}
\DeclareMathSymbol{\nub}{\mathord}{grb}{"17}
\DeclareMathSymbol{\rhob}{\mathord}{grb}{"1A}
\DeclareMathSymbol{\sigmab}{\mathord}{grb}{"1B}
\DeclareMathSymbol{\taub}{\mathord}{grb}{"1C}
\DeclareMathSymbol{\phib}{\mathord}{grb}{"1E}
\DeclareMathSymbol{\chib}{\mathord}{grb}{"1F}
\DeclareMathSymbol{\psib}{\mathord}{grb}{"20}
\DeclareMathSymbol{\omegab}{\mathord}{grb}{"21}
\DeclareMathSymbol{\epsilonb}{\mathord}{grb}{"22}
\DeclareMathSymbol{\varphib}{\mathord}{grb}{"27}
\begin{document}




\title{Joint DOA and Polarization Estimation \\with Sparsely Distributed and Spatially \\Non-Collocating Dipole/Loop Triads}


\author{
\authorblockN{Xin YUAN}\\
\authorblockA{Department of Electrical and Computer Engineering, \\
Duke University, Durham, NC, 27708 \\
xin.yuan@duke.edu}
}

\maketitle

\begin{abstract}
This paper introduces an ESPRIT-based algorithm to estimate the directions-of-arrival and polarizations for multiple sources.
The investigated algorithm is based on new sparse array geometries, which are composed of three non-collocating dipole triads or three non-collocating loop triads.
Both the inter-triad spacings and the inter-sensor spacings in the same triad can be far larger than a half-wavelength of the incident sources.
By adopting the ESPRIT algorithm, the eigenvalues of the data-correlation matrix offer the fine but ambiguous estimates of the direction-cosines for each source, and the eigenvectors provide the estimates of each source's steering vector.
Based on the constrained array geometries, the fine and unambiguous estimates of directions-of-arrival and polarizations are obtained.
Simulation results verify the efficacy of the investigated approach and also verify the aperture extension property of the proposed array geometries.
\end{abstract}


\section{Introduction}
Polarization is a property of the electromagnetic wave and it is intrinsically associated with the electric field and magnetic field of the wave \cite{Naidu01}.
Expressed in the Cartesian coordinate system, the electric field has three components along each axis, and the magnetic filed also has three components along each axis \cite{SwindlehurstAPT1293,WeissSPT0593,KrimSPM0796}.
A short dipole oriented along the axis in the Cartesian coordinate system is used to measure corresponding component of the electric field, and a small loop
oriented along the axis in the Cartesian coordinate system is used to measure corresponding component in the magnetic field \cite{LiAPT0393,HurtadoSPM0109}.
Two orthogonally-collocated dipoles will form a dipole-pair (a.k.a cross dipoles) \cite{LiAPT0991,
LiAPT1091,
ChengSPT1194,
GongSP0509,
HuaAPT0393}, and two orthogonally-collocated dipoles will form a loop-pair \cite{WongKT_APM0205,YuanAPT0512,
MirSPT0107}.
In addition, the dipole-loop pair is investigated in \cite{LiAPT0496,YuanAPT0512}.
Since both the electric field and magnetic field have three components in the Cartesian coordinate system, a dipole-triad 
\cite{WongKT_AEST0401,AuYeungAEST0109,YuanSJ0612,XuIET-MAP0612} comprises three orthogonally-collocated dipoles which are used to measure the three components of the signal's electric field, and
a loop-triad \cite{WongKT_AEST0401,AuYeungAEST0109,YuanSJ0612,LuoEURASIP0512} comprises three orthogonally-collocated loops which are used to measure the three components of the signal's magnetic field.
When the dipole triad and the loop triad are collocated at a point geometry in space, an electromagnetic vector-sensor \cite{LiAPT0393,NehoraiSPT0294,WongKT_APT0500,
WongKT_APT0800,
Zoltowski1SPT0800,
Zoltowski2SPT0800} will be constructed.
Advantages of these dipole/loop pair, triad, and electromagnetic vector-sensor are that they can resolve both the polarization and the direction-of-arrival (DOA) differences of the incident sources \cite{LiAPT0393,NehoraiSPT0294}.
Because all antennas in the vector-sensor are collocated, they will have the same frequency-dependence.
Therefore, the array-manifold of the vector-sensor is independent of the signal's frequency-spectrum \cite{WongKT_APT1097}.
The electromagnetic vector-sensor (array) has been investigated extensively for direction finding and polarization estimation with different algorithms:
for example: the Estimation of Signal Parameters Via Rotational Invariance Techniques (ESPRIT) \cite{RoyASSPT0789} based algorithms \cite{LiAPT0393,WongKT_APT1097,YuanSPT0312,HoSPT1097,HoSPT1099,
WongKT_APT0500,Zoltowski1SPT0800,
Zoltowski2SPT0800,XuJCIC0504,QiIET-MAP0411,JiJASP06},
the beamforming algorithms \cite{WongKT_AEST0101,XuIET-RSN0607},
the MUSIC based algorithms \cite{WongKT_APT0800,XuJCIC0504,RahamimSPT1104,MironSPT0406,LeBihanSPT0907},
the Root-MUSIC based algorithms \cite{WongKT_AWPL04,XuIET-RSN0607},
the maximum likelihood algorithms \cite{LeeSPT0294},
the quaternion based algorithms \cite{MironSPT0406,LeBihanSPT0907,GongSP0411,GongAEST0711,LeBihanSP0704},
the two-fold mode algorithms \cite{GongSP0509},
the parallel factor based algorithms \cite{GongIET-SP0411},
and
the propagator methods \cite{HeIET-RSN0509,HeDSP0509,LiuJEIT1010,HeAEST0110,LiuEURASIP2011}.
Theoretical bounds are analyzed in \cite{NehoraiSPT0294,FriedlanderSPT0794,HochwaldSPT0895,YuanSPT0312,HurtadoAEST0407,NordeboSPT1006}.
Source tracking with the electromagnetic vector-sensor is investigated in \cite{KoAEST0702,NehoraiSPT1099,YuanICASSP2012}.
The beampattern of the electromagnetic vector-sensor is studied in \cite{NehoraiSPT0399,XiaoSPT0209}.
The linear dependence of the steering vectors associated with these vector-sensors is investigated in \cite{HoSP95,HochwaldSPT0196,TanSPT1296,HoAPT1198,XuAEST1008,LiEL0707}.
Disadvantage of these collocated sensors is also obvious; the mutual coupling across the collocated antennas will increase the hardware cost of the vector-sensors and also degrade the performance \cite{WongKT_SPT0111,YuanSSP2011}.

In order to overcome the mutual coupling problem, distributed electromagnetic vector sensors have been investigated in
\cite{SeeSPA03,SeeSSP03,LoMonteWDDC07,LoMonteICEAA07,SunCISP09,SunCISP10,SunAPSEC10} recently.
Reference \cite{WongKT_SPT0111} proposed a non-collocating electromagnetic vector sensor by spatially spacing the three dipoles and three loops in the vector-sensor, and this array geometry remain the vector-cross-product direction finding algorithm, which has been investigated extensively for the collocated electromagnetic vector-sensor.
Advantages of the non-collocating electromagnetic vector sensor include that \cite{WongKT_SPT0111}:
1) the mutual coupling is effectively reduced since the inter-sensor spacings are far larger than a half-wavelength,
2) the angular resolution is enhanced because of the array aperture extension, and
3) the vector-cross-product direction finding approach is remained for fast DOA estimation.

However, in practical applications, the responses of electric field and electric field vary from each other \cite{LiAPT0393}. Hence only dipoles or loops are a better choice to form a polarized sensor array.
It follows that the vector-cross-product direction finding approach can not be used.
This paper proposes sparse arrays composed of non-collocating dipole triads or non-collocating loop triads.
From \cite{WongKT_AEST0401,YuanSJ0612}, we know that one a single dipole/loop triad suffices for direction finding and polarization estimation.
However, when the three dipoles/loops in the triad are spatially spread in space, they can not offer the closed-form direction-of-arrival and polarization estimation because of the inter-sensor phase factors, especially when the inter-sensor spacings are far larger than a half wavelength.
In this work, we synergize the ESPRIT \cite{RoyASSPT0789} algorithm with the non-collocating dipole/loop triads.
Sparse arrays composed of three non-collocating dipole triads or three non-collocating loop triads are proposed to estimate the DOAs and polarizations of multiple sources.
Both the inter-triad spacings and the inter-sensor spacings in the same triad can be far larger than a half-wavelength.

\subsection{Relative Research and New Contributions}
Recent research on the directions-of-arrival and polarizations estimation use various polarized antenna array configurations.
The {\em collocated} six-component electromagnetic vector sensor is used in \cite{YuanSPT0312,YuanICASSP2012}.
Reference \cite{YuanAPT0512} focus on the polarization estimation with different compositions of antenna pairs, which can be collocated or non-collocated.
Reference \cite{YuanSJ0612} investigates various {\em collocated} dipole/loop triads for direction-finding and polarization estimation.
Diversely {\em collocated} dipole and/or loop quads for direction-finding and polarization estimation are exploited in \cite{YuanAWPL2012}
Different from the aforementioned research, the present paper utilize the {\em spatially spread} non-collocating dipole/loop triads for multiple sources direction-finding and polarization estimation.
Contributions of this work are summarized as follows:
a) the mutual coupling across the antennas is reduced,
b) the angular resolution is enhanced,
c) the hardware cost is reduced,
d) a novel ESPRIT-based algorithm is derived based on the sparse array geometries,
e) both the eigenvectors and the eigenvalues are utilized during the estimation, and
f) a brief pair algorithm is proposed under the multiple-source scenario in the estimation procedure.

\subsection{Organization of This Paper}
The remainder of this paper is organized as follows:
Section \ref{Sec:geo} provides the array geometry used in this work.
Section \ref{Sec:ag} presents the algorithm to derive the closed-form estimation of arriving angles and polarizations in the multiple sources scenario.
Section \ref{Sec:sim} shows the simulation results to verify the performance of the proposed algorithm.
Section \ref{Sec:con} concludes the whole paper.

\section{Array Geometry}
\label{Sec:geo}
\begin{subfigures}
\begin{figure}[htbp]
\centering
\begin{minipage}{6in}
\includegraphics[scale=0.3]{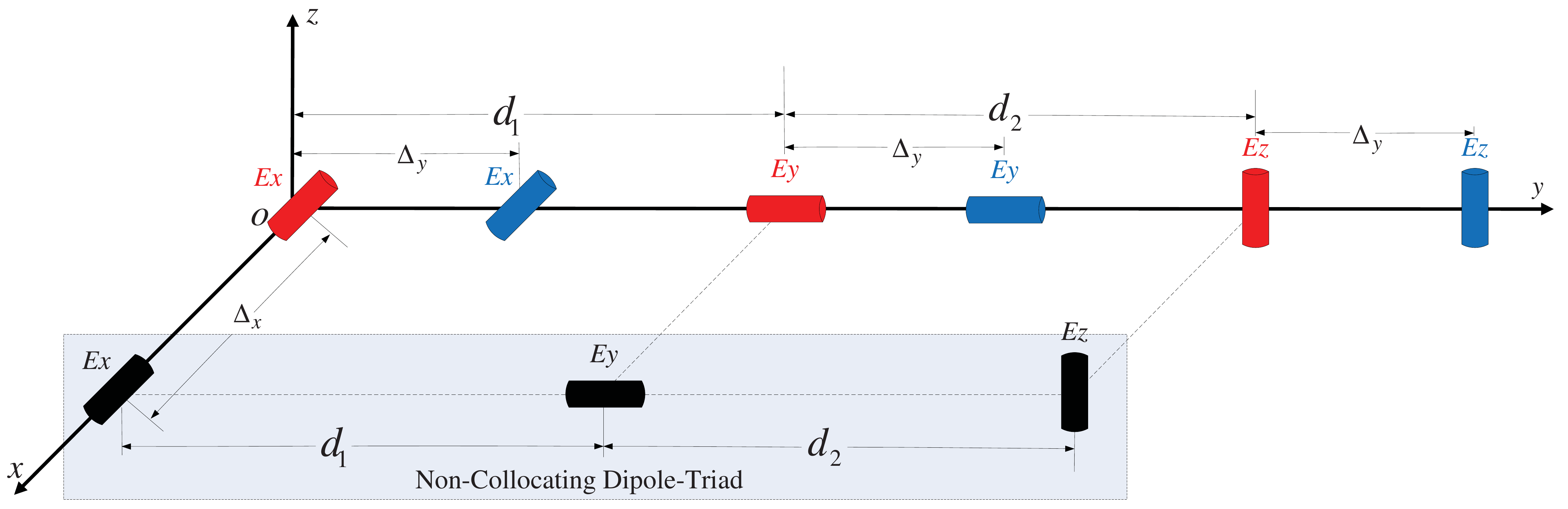}
\caption{Array geometry composed of three spatially spread non-collocating dipole triads. The inter-dipole spacings among the dipoles are all beyond $\lambda/2$. The three orthogonally oriented dipoles $\{E_x,E_y,E_z\}$ in the same color form one non-collocating dipole-triad.}
\label{DT-SS}
\end{minipage}
    \hfill
\begin{minipage}{6in}
\includegraphics[scale=0.3]{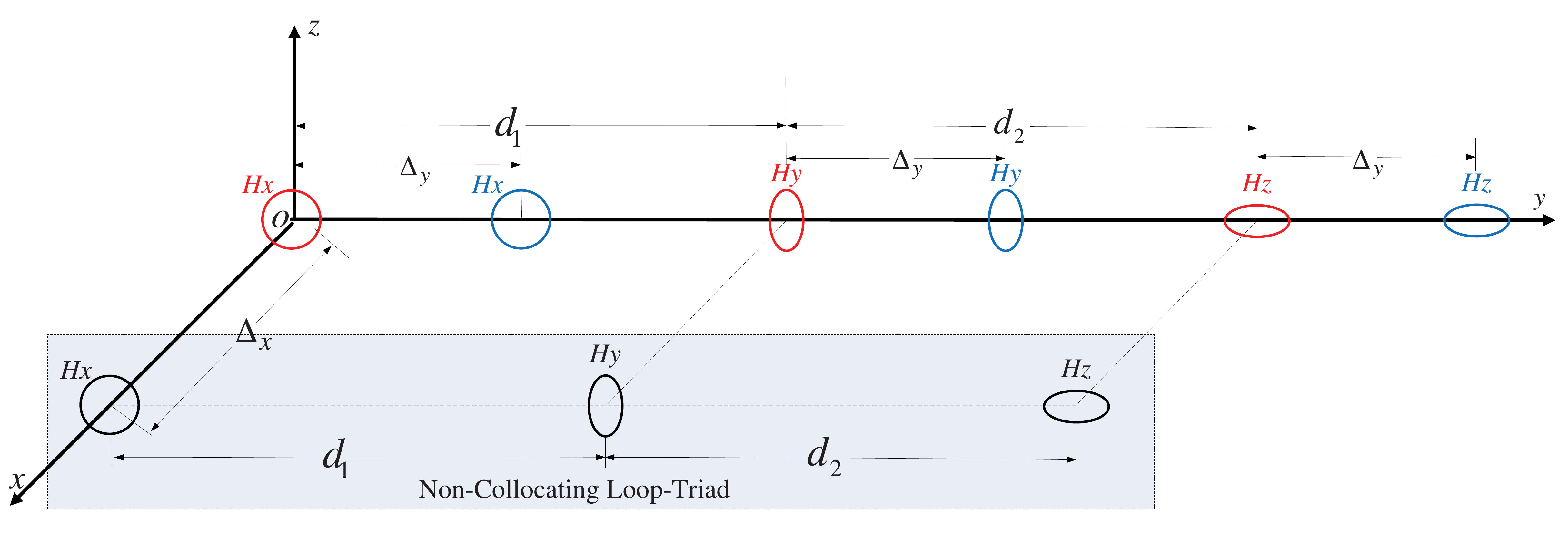}
\caption{Array geometry composed of three spatially spread non-collocating loop triads. The inter-loop spacings among the loops are all beyond $\lambda/2$.
The three orthogonally oriented loops $\{H_x,H_y,H_z\}$ in the same color form one non-collocating loop-triad.}
\label{LT-SS}
\end{minipage}
\label{DLDT-SS}
\end{figure}
\end{subfigures}

Figure \ref{DLDT-SS} depicts the array-geometries investigated in this paper.
The array composed of three non-collocating dipole triads is demonstrated in Figure \ref{DT-SS}, and the array composed of three non-collocating loop triads is demonstrated in Figure \ref{LT-SS}.
The three orthogonally oriented dipoles/loops in each dipole/loop triad are displaced on the $y$-axis with the distance between $e_x/h_x$ and $e_y/h_y$ equaling to $d_1$, and the distance between $e_y/h_y$ and $e_z/h_z$ equaling to $d_2$.
Another dipole/loop triad is also displaced on the $y$-axis with a distance $\Delta_y$ to the first triad.
A third dipole/loop triad is displaced parallel the $y$-axis with a distance $\Delta_x$ to the first triad.
Please note all the distances among the dipoles/loops can be far larger than a half-wavelength, $\lambda/2$, where $\lambda$ denotes the wavelength of the signal.
The following constrains are required to form the array:
\begin{eqnarray}
d_1 &=& m_1 \Delta_y, \hspace{0.1in} \mbox{where} \hspace{0.08in}m_1  \hspace{0.08in}\mbox{is a positive integer}, m_1=1,2,\cdots. \label{eq:d1Deltay}\\
d_2 &=& m_2 \Delta_y, \hspace{0.1in} \mbox{where}\hspace{0.08in} m_2  \hspace{0.08in}\mbox{is a positive integer}, m_2=1,2,\cdots. \label{eq:d2Deltay}
\end{eqnarray}
$d_1$ can be the same as or different from $d_2$, which means $m_1=m_2$ or $m_1\neq m_2$.
This constrain is critical to the proposed ESPRIT-based algorithm in the following section.

In a multiple-source scenario with $K$ sources, the responses of the dipoles along each axis for the $k$th signal are \cite{NehoraiSPT0294,LiAPT0393}:
\begin{eqnarray}
{\bf e}_k &\stackrel{\rm def}{=} & \left[\begin{array}{l}   e_{x,k}  \\
                          e_{y,k}  \\
                          e_{z,k}  \end{array}\right]
\hspace{0.08in}  \stackrel{\rm def}{=} \hspace{0.08in}
\left[\begin{array}{c}
                    \cos\theta_{1,k}\sin\theta_{2,k}\sin\theta_{3,k} e^{j\theta_{4,k}}- \sin\theta_{1,k}\cos\theta_{3,k} \\
\sin\theta_{1,k}\sin\theta_{2,k}\sin\theta_{3,k} e^{j\theta_{4,k}}+ \cos\theta_{1,k}\cos\theta_{3,k}\\
-\cos\theta_{2,k}\sin\theta_{3,k} e^{j\theta_{4,k}}
                   \end{array} \right], \label{eq:aDT}
\end{eqnarray}
where $\{\theta_{1,k}\in[0,2\pi), \theta_{2,k}\in[-\pi/2,\pi/2]\}$ are the azimuth-angle and elevation-angle of the $k$th source,
and $\{\theta_{3,k}\in[0,\pi/2],\theta_{4,k}\in[-\pi,\pi)\}$ denote the auxiliary polarization angle and polarization phase difference of the $k$th incident signal respectively (equating to $\{\gamma,\eta\}$ in \cite{LiAPT0393}).
The responses of the loops along each axis for the $k$th signal are \cite{NehoraiSPT0294,LiAPT0393}:
\begin{eqnarray}
{\bf h}_k &  \stackrel{\rm def}{=} &
 \left[\begin{array}{l}   h_{x,k}  \\
                          h_{y,k}  \\
                          h_{z,k} \end{array}\right]
\hspace{0.08in}  \stackrel{\rm def}{=} \hspace{0.08in}
\left[\begin{array}{c}
-\sin\theta_{1,k}\sin\theta_{3,k} e^{j\theta_{4,k}}-\cos\theta_{1,k}\sin\theta_{2,k}\cos\theta_{3,k}\\
\cos\theta_{1,k}\sin\theta_{3,k} e^{j\theta_{4,k}}- \sin\theta_{1,k}\sin\theta_{2,k}\cos\theta_{3,k}\\
\cos\theta_{2,k}\cos\theta_{3,k}
                   \end{array} \right].\label{eq:aLT}
\end{eqnarray}

Since the dipoles or loops in Figure \ref{DLDT-SS} are spatially-spread, the inter-sensor phase factors will be introduced in the array-manifold.
The array-manifold of the non-collocating dipole triad in Figure \ref{DT-SS} corresponding to $k$th source is:
\begin{eqnarray}
{\bf a}_{{\rm nc},k}&\stackrel{\rm def}{=}&
\left[\begin{array}{r}
e_{x,k} \\
e^{-j\frac{2\pi u_{y,k}}{\lambda_k}d_1 }e_{y,k}  \\
e^{-j\frac{2\pi u_{y,k}}{\lambda_k}(d_1+d_2)}e_{z,k}    \\
\end{array}\right], \label{eq:aDTss}
\end{eqnarray}
where $u_{y,k}=\cos\theta_{2,k}\sin\theta_{1,k}$ is the direction-cosine of the $k$th source align to $y$-axis.
The array-manifold of the non-collocating loop triad in Figure \ref{LT-SS} corresponding to $k$th source is:
\begin{eqnarray}
{\bf a}_{{\rm nc},k}&\stackrel{\rm def}{=}&
\left[\begin{array}{r}
h_{x,k}   \\
e^{-j\frac{2\pi u_{y,k}}{\lambda_k}d_1 } h_{y,k}  \\
e^{-j\frac{2\pi u_{y,k}}{\lambda_k}(d_1+d_2)}h_{z,k}   \\
\end{array}\right] \label{eq:aLTss}
\end{eqnarray}

The array-manifold of the
demonstrated array in Figure \ref{DLDT-SS} for $k$th source is thus a $9 \times1$ vector:
\begin{eqnarray} \label{eq:bfa}
{\bf a}_k&=& {\bf a}_{{\rm nc},k}\otimes \left[\begin{array}{l}
1\\
q_{y,k} \\
q_{x,k}
\end{array}\right]
\end{eqnarray}
where $\otimes$ denotes the Kronecker-product operator, and
\begin{eqnarray}
q_{y,k} &=& e^{-j\frac{2\pi}{\lambda}\Delta_y u_{y,k}},\\
q_{x,k} &=& e^{-j\frac{2\pi}{\lambda}\Delta_x u_{x,k}},
\end{eqnarray}
with $u_{x,k}\stackrel{\rm def}{=}\cos\theta_{2,k}\cos\theta_{1,k}$ denoting the direction-cosine of the $k$th source along $x$-axis.

The following will derive an ESPRIT-based algorithm to estimate the directions-of-arrival and polarizations of multiple incident sources based on the array geometries in Figure \ref{DLDT-SS}.

\section{Algorithm Derivation}
\label{Sec:ag}
In a $K$-source scenario, the data set measured at time $t$ by the array in Section \ref{Sec:geo} is:
\begin{eqnarray} \label{eq:bfyt}
{\bf y}(t) &=& \sum^K_{k=1}{\bf a}_k s_k (t) + {\bf n}(t),
\end{eqnarray}
where ${\bf a}_k$ is the steering-vector of the $k$th source as shown in (\ref{eq:bfa}), ${\bf n}(t)$ is the zero-mean circularly symmetric additive
white Gaussian noise,
and $s_k(t)$ is the $k$th signal.

Decompose (\ref{eq:bfyt}) into three different parts:
\begin{eqnarray} \label{eq:y123t}
{\bf y}(t) &\stackrel{\rm def}{=}&\left[\begin{array}{l}
{\bf y}_1(t) \\
{\bf y}_2(t) \\
{\bf y}_3(t)
\end{array}\right] =\left[\begin{array}{l}
\sum^K_{k=1}{\bf a}_{{\rm nc},k} s_k (t) + {\bf n}_1(t) \\
\sum^K_{k=1}{\bf a}_{{\rm nc},k}q_{y,k} s_k (t) + {\bf n}_2(t) \\
\sum^K_{k=1}{\bf a}_{{\rm nc},k}q_{x,k} s_k (t) + {\bf n}_3(t)
\end{array}\right]
\end{eqnarray}

\subsection{Basic Principle Underlying the Algorithm}
The main idea of the algorithm investigated in this paper is to creatively use the ESPRIT algorithm in the polarized antenna arrays demonstrated in Figure \ref{DLDT-SS}. Unlike the general scalar antenna array, the proposed antenna arrays a) are polarized, b) are sparse arrays, and c) are composed of non-collocating dipole/loop triads.
Similar to the unpolarized uniform antenna array, the two dimensional ESPRIT algorithm is used.
Different from the uniformly scalar sensor array,
1) both the eigenvalues and the eigenvectors of the data-correlation matrix will be used,
2) the eigenvalues will present the fine estimates of the direction-cosines,
3) the steering vectors of the sources are estimated from the eigenvectors,
4) the fine estimates of the direction-cosines along $y$-axis will be used to eliminate the inter-sensor phase factors in the steering vectors of the sources collected by the non-collocating dipole/loop triads,
5) the coarse estimates of directions-of-arrival and polarizations are estimated from the steering vectors derived from the eigenvectors and so they are automatically associated with each other for each source,
6) after the coarse estimates of directions-of-arrival are derived, the coarse estimates of the direction-cosines along $x$-axis are estimated,
7) the eigenvalues used to derive the fine estimates of the direction-cosines along $x$-axis are paired with the coarse estimates, and the proposed pair algorithm is very brief with a low computation workload,
8) the coarse estimates are used to disambiguate the fine estimates to obtain the final, the fine and unambiguous, estimates of the direction-cosines,
and lastly
9) the directions-of-arrival and polarizations for each source are derived.

\subsection{Adopt the ESPRIT Algorithm}
Consider there are $M$ time samples collected at time $t_1,t_2,\cdots,t_M$, from (\ref{eq:y123t})
\begin{eqnarray}
{\bf Y}&=& \left[{\bf y}(t_1), {\bf y}(t_2),\cdots {\bf y}(t_M)\right] \stackrel{\rm def}{=}\left[\begin{array}{l}
{\bf Y}_1 \\
{\bf Y}_2 \\
{\bf Y}_3
\end{array}\right]
\end{eqnarray}
Compute the data correlation matrix of ${\bf Y}$:
\begin{eqnarray} \label{eq:bfR}
{\bf R} &=& {\bf Y}{\bf Y}^H,
\end{eqnarray}
where $^H$ is the Hermitian operator.
The steering vectors corresponding to ${\bf R}$ in (\ref{eq:bfR}) are:
\begin{eqnarray}
{\bf A} &\stackrel{\rm def}{=}& \left[\begin{array}{c}
{\bf A}_1 \\
{\bf A}_2 \\
{\bf A}_3
\end{array}\right] = \left[\begin{array}{l}
{\bf A}_1 \\
{\bf A}_1 {\bf \Phi}_y \\
{\bf A}_2 {\bf \Phi}_x
\end{array}\right],
\end{eqnarray}
where ${\bf A}_1\stackrel{\rm def}{=}\left[{\bf a}_{{\rm nc},1}, {\bf a}_{{\rm nc},2}, \cdots, {\bf a}_{{\rm nc},K}\right]$,
and ${\bf \Phi}_y ={\rm diag}\left[e^{-j\frac{2\pi}{\lambda_1}u_{y,1} \Delta_y}, e^{-j\frac{2\pi}{\lambda_2}u_{x,2} \Delta_y}, \cdots, e^{-j\frac{2\pi}{\lambda_K}u_{x,K} \Delta_y}\right]$,
${\bf \Phi}_x ={\rm diag}\left[e^{-j\frac{2\pi}{\lambda_1}u_{x,1} \Delta_x}, e^{-j\frac{2\pi}{\lambda_2}u_{x,2} \Delta_x}, \cdots, e^{-j\frac{2\pi}{\lambda_K}u_{x,K} \Delta_x}\right]$.

Perform the eigen-decomposition of the covariance matrix ${\bf R}$:
\begin{eqnarray}
{\bf R}  &=& {\bf E}_s {\Lambda}_s  {\bf E}^H_s + {\bf E}_n {\Lambda}_n  {\bf E}^H_n,
\end{eqnarray}
where ${\bf E}_s$ is the signal subspace composed of the eigen-vectors associated with the $K$ largest eigen-values.
Partition the $9\times K$ signal subspace ${\bf E}_s$ into three $3\times K$ sub-matrices, ${\bf E}_{s,1}, {\bf E}_{s,2},{\bf E}_{s,3}$,
where ${\bf E}_{s,1}$ is composed of the top 3 rows, ${\bf E}_{s,2}$ is composed of the middle 3 rows, and ${\bf E}_{s,3}$ is composed of the bottom 3 rows.
In the {\em noiseless} case, ${\bf E}_{s,1}, {\bf E}_{s,2}$ and ${\bf E}_{s,3}$ are inter-related with each other by:
\begin{eqnarray}
{\bf E}_{s,2} &=& {\bf E}_{s,1} {\bf \Phi}_y, \\
{\bf E}_{s,3} &=& {\bf E}_{s,1} {\bf \Phi}_x,
\end{eqnarray}
In the {\em noisy} case, ${\bf \Phi}_y$ and ${\bf \Phi}_x$ can be estimated by \cite{RoyASSPT0789}:
\begin{eqnarray}
{\hat{\bf \Phi}}_y &=& \left({\bf E}^H_{s,1} {\bf E}_{s,1}\right)^{-1} {\bf E}^H_{s,1} {\bf E}_{s,2}, \\
{\hat{\bf \Phi}}_x &=& \left({\bf E}^H_{s,1} {\bf E}_{s,1}\right)^{-1} {\bf E}^H_{s,1} {\bf E}_{s,3}.
\end{eqnarray}

There exists a unique $K\times K$ nonsingular matrix ${\bf T}$ such that \cite{RoyASSPT0789}:
\begin{eqnarray}
{\bf E}_{s,1} &=& {\bf A}_1 {\bf T},\\
{\bf E}_{s,2} &=& {\bf A}_2 {\bf T} = {\bf A}_1 {\bf \Phi}_y {\bf T}, \\
{\bf E}_{s,3} &=& {\bf A}_3 {\bf T} = {\bf A}_1 {\bf \Phi}_x {\bf T}.
\end{eqnarray}
In the {\em noisy} case, this ${\bf T}$ can be estimated by performing the eigen-decomposition of ${\hat{\bf \Phi}}_y$ and ${\hat{\bf \Phi}}_x$.
${\hat{\bf T}}$ is composed of the eigenvectors, and
\begin{eqnarray}
{\bf D}_y={\rm diag} \left[\sigma_{1,y},\sigma_{2,y},\cdots, \sigma_{K,y}\right],\label{eq:Deltay}\\
{\bf D}_x={\rm diag} \left[\sigma_{1,x},\sigma_{2,x},\cdots, \sigma_{K,x}\right], \label{eq:Deltax}
\end{eqnarray}
comprise the eigenvalues.
${\bf D}_y$ will offer the {\em fine but ambiguous} estimates of the sources' direction-cosines along the $y$-axis,
and ${\bf D}_x$ will offer the {\em fine but ambiguous} estimates of the sources' direction-cosines along the $x$-axis.
However, since the eigen-decomposition operations of ${\hat{\bf \Phi}}_y,{\hat{\bf \Phi}}_x$ are independent with each other, the direction-cosines estimates
in ${\bf D}_y, {\bf D}_x$ should be paired.
Furthermore, these estimates need to be disambiguated.
The following will show how.

\subsection{Direction-of-Arrival Estimation}
Now consider ${\bf D}_y$ and the eigen-decomposition of ${\hat{\bf \Phi}}_y$,
The steering vectors of the sources can be estimated by \cite{WongKT_APT1097}:
\begin{eqnarray} \label{eq:hatA1}
\hat{\bf A}_1 &=& {\bf E}_{s,1}  {\hat{\bf T}}^{-1} + {\bf E}_{s,2}  {\hat{\bf T}}^{-1} {\bf D}_y^{-1}
= \left[\hat{\bf a}_{{\rm nc},1}, \hat{\bf a}_{{\rm nc},2}, \cdots, \hat{\bf a}_{{\rm nc},K}\right].
\end{eqnarray}
and the {\em fine but ambiguous} estimates of the $k$th source's direction-cosine along the $y$-axis:
\begin{eqnarray} \label{eq:vfine}
\hat{u}_{y,k}^{\rm fine} &=& -\frac{\lambda_k}{2\pi \Delta_y}\angle \sigma_{k,y}, \hspace{0.1in} \forall k=1,\cdots,K.
\end{eqnarray}
where $\angle$ denotes the complex angle of the ensuing number.
Since $\Delta_y\ge \lambda_k/2$, there exists a unique integer $n^{\circ}_{y,k}$ that:
\begin{eqnarray}
{\hat u}_{y,k} &=& {\hat u}^{\rm fine}_{y,k} + n^{\circ}_{y,k} \frac{\lambda_k}{\Delta_y},  \label{eq:uyk}\\
 \frac{2\pi \Delta_y}{\lambda_k}&=& 2n^{\circ}_{y,k}\pi- \angle \sigma_{k,y}.
\end{eqnarray}
It follows that:
\begin{eqnarray}
\frac{2\pi d_2}{\lambda_k}&=& \frac{d_2}{\Delta_y} \frac{2\pi \Delta_y}{\lambda_k} = \frac{d_2}{\Delta_y}\left(2n^{\circ}_{y,k}\pi- \angle \sigma_k\right),\\
\frac{2\pi (d_1+d_2)}{\lambda_k}&=& \frac{(d_1+d_2)}{\Delta_y} \frac{2\pi \Delta_y}{\lambda_k} = \frac{(d_1+d_2)}{\Delta_y}\left(2n^{\circ}_{y,k}\pi- \angle \sigma_k\right).
\end{eqnarray}
Recall (\ref{eq:d1Deltay})-(\ref{eq:d2Deltay}) that $d_1=m_1\Delta_y, d_2=m_2 \Delta_y$, we can obtain
\begin{eqnarray}
\frac{2\pi d_2}{\lambda_k}&=& m_2\left(2n^{\circ}_{y,k}\pi- \angle \sigma_k\right)=2m_2n^{\circ}_{y,k}\pi - m_2\angle \sigma_{k,x}, \label{eq:2pid1}\\
\frac{2\pi (d_1+d_2)}{\lambda_k}&=& 2(m_1+m_2)n^{\circ}_{y,k}\pi - (m_1+m_2)\angle \sigma_{k,x}. \label{eq:2pid1d2}
\end{eqnarray}

Note that $\hat{\bf A}_1$ will offer the estimates of the the steering vectors.
We can obtain from $\hat{\bf A}_1 $ that $\hat{\bf a}_{{\rm nc},k}= c{\bf a}_{{\rm nc},k}$, where $c$ is an unknown complex number.
From $\hat{\bf a}_{{\rm nc},k}= c{\bf a}_{{\rm nc},k}$ and (\ref{eq:2pid1})-(\ref{eq:2pid1d2}), we can define:
\begin{eqnarray}
{\bf d}&=& \left[\begin{array}{l}
\frac{[{\hat{\bf a}}_{{\rm nc},k}]_1}{[{\hat{\bf a}}_{{\rm nc},k}]_3} e^{- j(m_1+m_2)\angle \sigma_{k,y}}\\
\frac{[{\hat{\bf a}}_{{\rm nc},k}]_2}{[{\hat{\bf a}}_{{\rm nc},k}]_3} e^{-jm_2\angle \sigma_{k,y}}
\end{array}\right] = \left\{\begin{array}{cc}\left[\frac{e_x}{e_z},  \frac{e_y}{e_z}\right]^T & \mbox{for the array in Figure \ref{DT-SS}},\\
\left[\frac{h_x}{h_z},  \frac{h_y}{h_z}\right]^T & \mbox{for the array in Figure \ref{LT-SS}}.
\end{array}\right.
\end{eqnarray}
where $^T$ denotes the transposition.

From the equations derived in \cite{YuanSPT0312,YuanSJ0612}, we can get:
\begin{eqnarray}
\hat{\theta}^{\rm coarse}_{1,k} &=& \left\{\begin{array}{ll}\tan^{-1}\left(\frac{-{\mathfrak Im}\{[{\bf d}]_1\}}{{\mathfrak Im}\{[{\bf d}]_2\}}\right), &{\mbox{if}}\hspace{0.04in}  ({\mathfrak Im}\{[{\bf d}]_2\}\sin\theta_{4,k})\ge0 \\
 \tan^{-1}\left(\frac{-{\mathfrak Im}\{[{\bf d}]_1\}}{{\mathfrak Im}\{[{\bf d}]_2\}}\right) +\pi, &{\mbox{if}} \hspace{0.04in}  ({\mathfrak Im}\{[{\bf d}]_2\}\sin\theta_{4,k})<0
\end{array}\right.  \label{eq:phidt}\\
\hat{\theta}^{\rm coarse}_{2,k} &=& \left\{\begin{array}{l}
\tan^{-1}\left(-{\mathfrak Re}\{[{\bf d}]_1\}\cos{\hat \theta^{\rm coarse}_{1,k}} - {\mathfrak Re}\{[{\bf d}]_2\}\sin{\hat \theta^{\rm coarse}_{1,k}}\right), \\
\hspace{0.5in}{\mbox{if}} \hspace{0.1in} \left({\mathfrak Re}\{[{\bf d}]_1\}\cos{\hat \theta^{\rm coarse}_{1,k}} + {\mathfrak Re}\{[{\bf d}]_2\}\sin{\hat \theta^{\rm coarse}_{1,k}}\right)\le0  \\
\tan^{-1}\left(-{\mathfrak Re}\{[{\bf d}]_1\}\cos{\hat \theta^{\rm coarse}_{1,k}} - {\mathfrak Re}\{[{\bf d}]_2\}\sin{\hat \theta^{\rm coarse}_{1,k}}\right)+\pi, \\
\hspace{0.5in}{\mbox{if}} \hspace{0.1in} \left({\mathfrak Re}\{[{\bf d}]_1\}\cos{\hat \theta^{\rm coarse}_{1,k}} + {\mathfrak Re}\{[{\bf d}]_2\}\sin{\hat \theta^{\rm coarse}_{1,k}}\right)>0 \end{array}\right. \\
\hat{\theta}^{\rm coarse}_{4,k} &=&  -\angle\left([{\bf d}]_1\sin{\hat \theta^{\rm coarse}_{1,k}}-[{\bf d}]_2\cos{\hat \theta^{\rm coarse}_{1,k}}\right) \label{eq:theta4}\\
\hat{\theta}^{\rm coarse}_{3,k} &=& \left\{\begin{array}{ll}
\cot^{-1}\left(\frac{{\mathfrak Im}\{[{\bf d}]_2\}\cos{\hat\theta}^{\rm coarse}_{2,k}}{\sin{\hat\theta_{4,k}}\cos{\hat\theta}^{\rm coarse}_{1,k}}\right), & \hspace{0.1in}\mbox{for the array in Figure \ref{DT-SS}};\\
\tan^{-1}\left(\frac{{\mathfrak Im}\{[{\bf d}]_2\}\cos{\hat\theta}^{\rm coarse}_{2,k}}{\sin{\hat\theta_{4,k}}\cos{\hat\theta}^{\rm coarse}_{1,k}}\right), & \hspace{0.1in}\mbox{for the array in Figure \ref{LT-SS}}.
\end{array} \right. \label{eq:theta3}
\end{eqnarray}
where ${\mathfrak Re}\{\hspace{0.03in}\}$ and ${\mathfrak Im}\{\hspace{0.03in}\}$ denote the real part and the imaginary part of the entry in $\{\hspace{0.03in}\}$, respectively.
Thus,
\begin{eqnarray}
{\hat u}^{\rm coarse}_{x,k}&=& \cos{\hat\theta}^{\rm coarse}_{2,k}\cos{\hat\theta}^{\rm coarse}_{1,k}, \label{eq:ucoarse}\\
{\hat u}^{\rm coarse}_{y,k}&=& \cos{\hat\theta}^{\rm coarse}_{2,k}\sin{\hat\theta}^{\rm coarse}_{1,k}, \label{eq:vcoarse}
\end{eqnarray}
and then the disambiguation method can be adopted to derive the final estimates of direction-cosines.
Using the coarse estimate of direction-cosine in (\ref{eq:vcoarse}) to disambiguate the fine estimate in
(\ref{eq:vfine}) by the method in \cite{Zoltowski1SPT0800,Zoltowski2SPT0800,WongKT_SPT0111}, we can determine $n^{\circ}_{y,k}$ in (\ref{eq:uyk}) and then derive the final, fine and unambiguous, estimate of direction-cosine ${\hat u}_{y,k}$.
For the details of this disambiguation, please refer to \cite{Zoltowski1SPT0800,Zoltowski2SPT0800,WongKT_SPT0111}.

The following problem is to get the final, fine and unambiguous, estimate of direction-cosine ${\hat u}_{x,k}$, and at first we need to pair the coarse estimate in (\ref{eq:ucoarse}) to the fine estimate in ${\bf D}_x$.
From ${\hat u}^{\rm coarse}_{x,k}$ in (\ref{eq:ucoarse}),
\begin{eqnarray}
\hat{q}^{\rm coarse}_{x,k} &=& e^{-j\frac{2\pi \Delta_x {\hat u}^{\rm coarse}_{x,k}}{\lambda_k}}.
\end{eqnarray}
On the other hand, from ${\bf D}_x$ in (\ref{eq:Deltax}),
\begin{eqnarray}
\hat{q}^{\rm fine}_{x,k} &=& \sigma_{k^{\prime}_0,x},
\end{eqnarray}
where
\begin{eqnarray}
k^{\prime}_0&=&\stackrel{\rm arg \hspace{0.05in} min}{k^{\prime}}\left|\hat{q}^{\rm coarse}_{x,k}-\sigma_{k^{\prime},x}\right|,\hspace{0.1in}\forall k^{\prime}=1,2,\cdots,K.
\end{eqnarray}
Then:
\begin{eqnarray}
\frac{2\pi \Delta_x}{\lambda_k}&=& 2n^{\circ}_{x,k}\pi- \angle \sigma_{k^{\prime}_0,x}, \\
{\hat u}_{x,k} &=& {\hat u}^{\rm fine}_{x,k} + n^{\circ}_{x,k} \frac{\lambda_k}{\Delta_x},
\end{eqnarray}
and this $ n^{\circ}_{x,k}$ can be determined by the coarse estimate of direction-cosine in (\ref{eq:ucoarse}).
Note that the above pair algorithm is a unique method developed for the proposed arrays and it has a low computation workload.

Lastly, after the unique $\{{\hat u}_{x,k}, {\hat u}_{y,k}\}$  has been obtained, the direction-of-arrival of $k$th incident source $\{\theta_{1,k},\theta_{2,k}\}$ can be estimated by \cite{WongKT_APT1097}:
\begin{eqnarray}
{\hat\theta}_{1,k} &=& \angle\left({\hat u}_{x,k}+j \hspace{0.03in}{\hat u}_{y,k}\right), \label{eq:hath1}\\
{\hat\theta}_{2,k} &=& \arccos\left(\sqrt{{{\hat u}_{y,k}}^2+{{\hat u}_{x,k}}^2}\right). \label{eq:hath2}
\end{eqnarray}
Then substituting ${\hat\theta}_{1,k},{\hat\theta}_{2,k}$ in (\ref{eq:hath1})-(\ref{eq:hath2}) into (\ref{eq:theta4})-(\ref{eq:theta3}) to replace ${\hat\theta}^{\rm coarse}_{1,k},{\hat\theta}^{\rm coarse}_{2,k}$, we can get the final estimates of the polarization parameters.

\section{Monte Carlo Simulation}
\label{Sec:sim}

The investigated algorithm's direction-finding efficacy and extended-aperture capability are demonstrated by Monte Carlo simulations.
The estimates use
$100$ temporal snapshots and $200$ independent
runs.
The root mean square error (RMSE) is utilized
as the performance measure. The RMSE for the direction-cosine is defined as:
\begin{eqnarray}
{\rm RMSE}&=& \sqrt{\frac{1}{200}\sum^{200}_{i=1}\left[({\hat u_x}^i - u_x)^2 + ({\hat u_y}^i - u_y)^2\right]}, \notag
\end{eqnarray}
where $\{{\hat u_x}^i,{\hat u_y}^i\}$ are the estimates of direction-cosines at $i$th run.
Figures \ref{RMSE-SNR2} plots the RMSEs of the direction-cosines versus signal-to-noise ratio (SNR) in a two-source scenario with sparse arrays proposed in Figure \ref{DT-SS}.
The two sources are with the digital frequencies $f_1 =0.0895, f_2=0.1685$.
The DOAs and polarizations of the two sources are set as:
$(\theta_{1,1},\theta_{2,1},\theta_{3,1},\theta_{4,1})=(30^{\circ},15^{\circ},45^{\circ},90^{\circ})$,
$(\theta_{1,2},\theta_{2,2},\theta_{3,2},\theta_{4,2})=(73^{\circ},43^{\circ},45^{\circ},-90^{\circ})$.
In the simulation, the inter-sensor spacing $d_1=d_2=8\lambda, \Delta_x=\Delta_y=4\lambda$, where $\lambda$ is the minimum wavelength of the sources.
Figures \ref{RMSE-SNR3} plots the RMSEs of the direction-cosines versus SNR in a three-source scenario.
The third source is with $f_3=0.2555$, $(\theta_{1,3},\theta_{2,3},\theta_{3,3},\theta_{4,3})=(150^{\circ},77^{\circ},45^{\circ},90^{\circ})$.

\begin{subfigures}
\begin{figure}[ht!]
\centering
\begin{minipage}{3in}
\includegraphics[height=6cm,width=8.0cm]{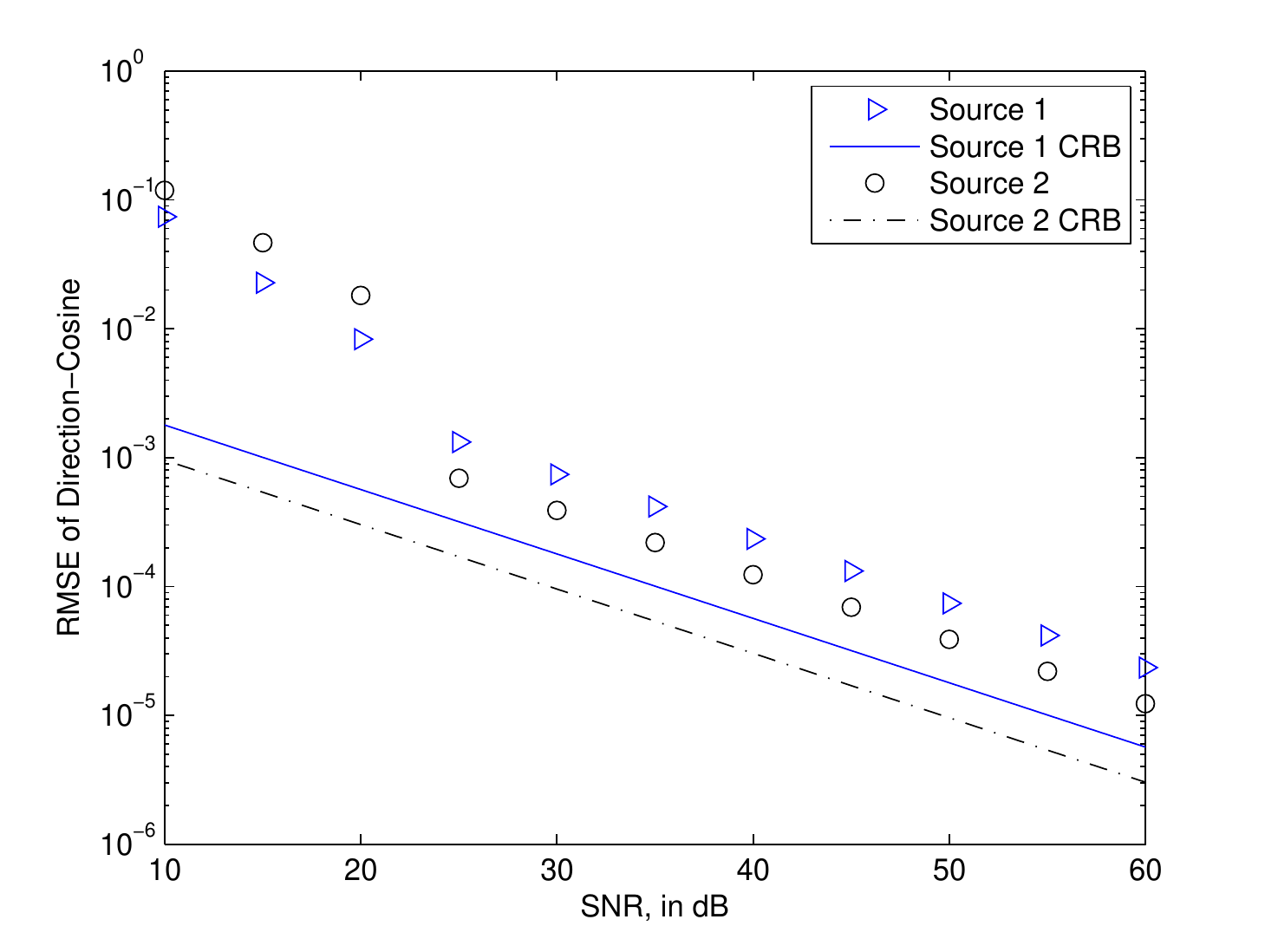}
\caption{Estimation RMSEs of direction-cosines versus SNR in a two-source scenario.}
\label{RMSE-SNR2}
\end{minipage}
    \hfill
\begin{minipage}{3in}
\includegraphics[height=6cm,width=8.0cm]{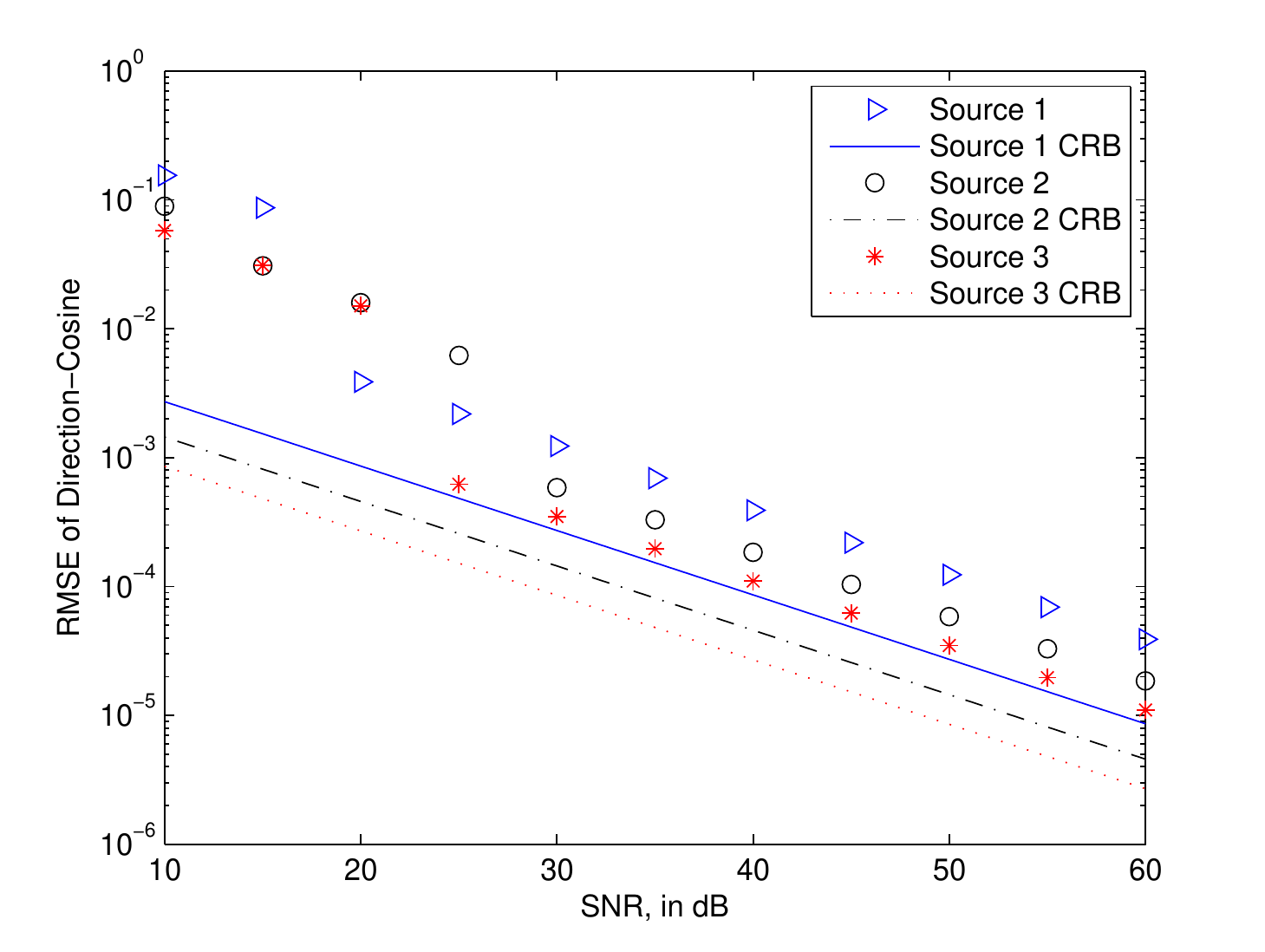}
\caption{Estimation RMSEs of direction-cosines versus SNR in a three-source scenario.}
\label{RMSE-SNR3}
\end{minipage}
\label{RMSE-SNR}
\end{figure}
\end{subfigures}

\begin{figure}[htbp]
\centering
\includegraphics[height=6cm,width=8.0cm]{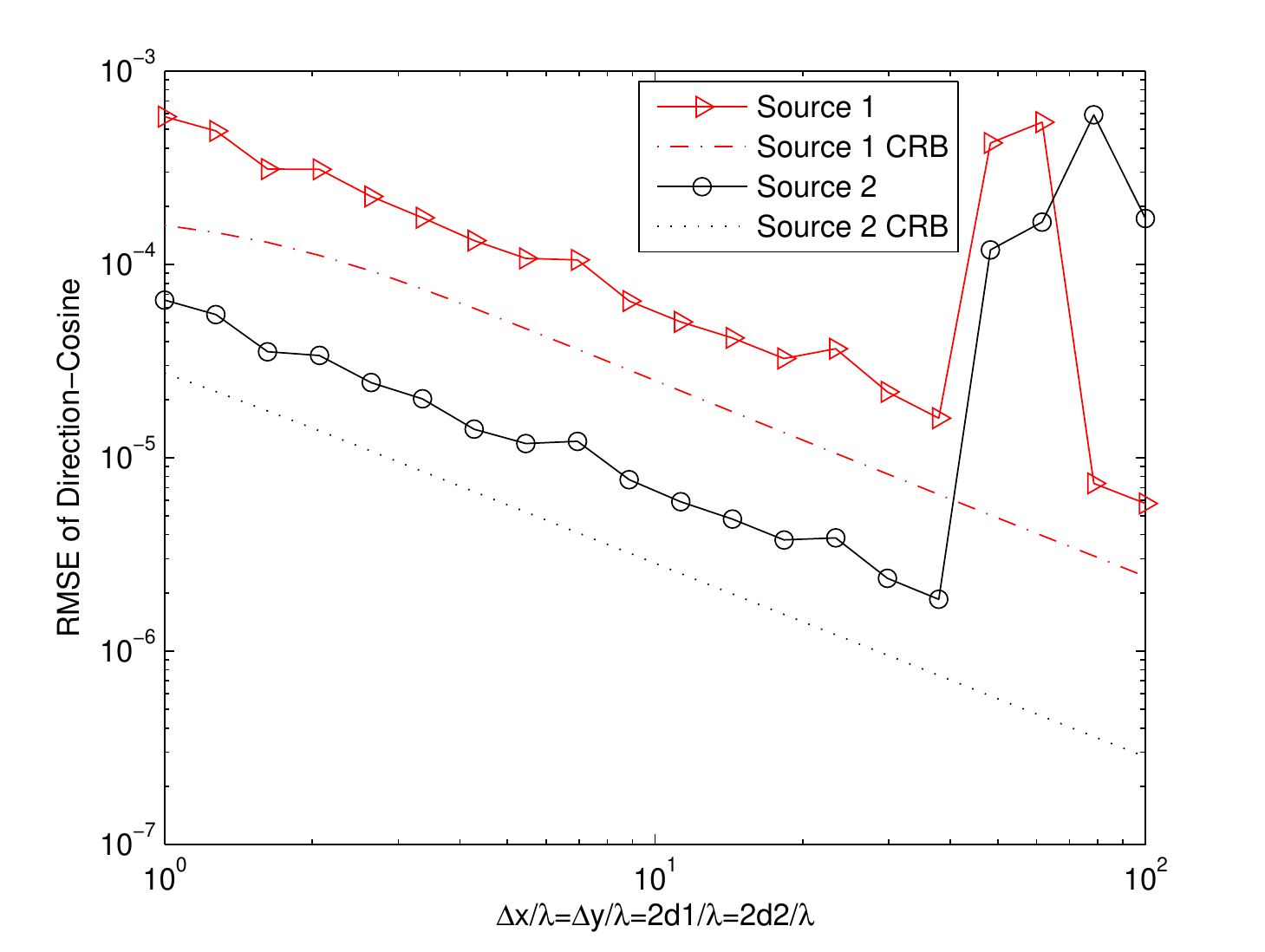}
\caption{The RMSEs of direction-cosines versus inter-sensor spacing $\frac{d_1}{\lambda}=\frac{\lambda}{d_2}=\frac{2\Delta_x}{\lambda}=\frac{2\Delta_y}{\lambda}$ in a two-source scenario, at SNR$=40$dB.}
\label{RMSE_kk}
\end{figure}

It is well known that the larger the array-aperture, the better the angular resolution.
In order to investigate the aperture extension property of the proposed array configurations, Figure \ref{RMSE_kk} plots the
RMSEs of direction-cosines versus inter-sensor spacings $\frac{d_1}{\lambda}=\frac{\lambda}{d_2}=\frac{2\Delta_x}{\lambda}=\frac{2\Delta_y}{\lambda}$ at SNR$=40$dB.
It can be seen that the RMSEs of direction-cosines estimated by the proposed algorithm decrease with the increase of inter-sensor spacings and they are close to the Cram\'{e}r-Rao bounds (CRB).
It is notable that there is a breakdown phenomenon in Figure \ref{RMSE_kk}. When the inter-sensor spacing $\frac{d_1}{\lambda}=\frac{\lambda}{d_2}=2\frac{\Delta_x}{\lambda}=2\frac{\Delta_y}{\lambda}$ is beyond a specific spacing point (about 40), the RMSEs of the final estimates will be the same as the coarse estimates. This is because the coarse estimates will identify the wrong estimation grid at the pre-set SNR and thus it can not be used to disambiguate the fine estimates. For the details of this breakdown phenomenon, please refer to \cite{Zoltowski1SPT0800,Zoltowski2SPT0800}.

\section{Conclusion}
\label{Sec:con}

A novel ESPRIT-based algorithm is investigated in this paper to estimate the directions-of-arrival and polarizations of multiple sources based on the proposed sparse arrays. Unlike the algorithms in \cite{Zoltowski1SPT0800,Zoltowski2SPT0800}, which is investigated for the collocated electromagnetic vector sensors, the sparse arrays studied in this work are composed of non-collocating dipole/loop triads.
The inter-sensor spacing in the arrays are far larger than a half-wavelength.
The mutual coupling across the antennas are thus reduced and the angular resolution is improved.

\small
\bibliographystyle{IEEEtran}


\end{document}